\documentclass[a4paper,11pt]{article}
\usepackage{jcappub}
\usepackage[utf8]{inputenc}
\usepackage{amsmath,mathtools}

\usepackage{tensor}
\usepackage{amssymb}
\usepackage{graphicx}
\usepackage{physics}
\usepackage{lipsum}  
\usepackage{bigints}
\usepackage{verbatim}
\usepackage[normalem]{ulem}
\usepackage[dvipsnames]{xcolor}
\newcommand{\leri}[1]{\left(#1\right)}

\newcommand{\act}{\frac{1}{2\kappa^2}\int d^3x\, \sqrt{-g}\;}
\newcommand{\pal}{\mathcal{R}}

\title{Slowly rotating charged BTZ black hole solutions in Palatini  Chern-Simons gravity}

\author[a]{Flavio Bombacigno,\note{Corresponding author.}}
\author[a,b]{Gonzalo J. Olmo,}
\author[c]{Emanuele Orazi}
\author[d]{Paulo J. Porf\'{i}rio}

\affiliation[a]{Instituto de Física Corpuscular (IFIC), CSIC‐Universitat de València, Spain}
\affiliation[b]{Universidade Federal do Cear\'a (UFC), Departamento de F\'isica,\\ Campus do Pici, Fortaleza - CE, C.P. 6030, 60455-760 - Brazil.}
\affiliation[c]{Escola de Ci\^{e}ncia e Tecnologia and International Institute of Physics,
Federal University of Rio Grande do Norte, Campus Universit\'{a}rio-Lagoa Nova, Natal-RN 59078-970, Brazil.}
\affiliation[d]{Departamento de F\'{\i}sica, Universidade Federal da Para\'{\i}ba,\\
Caixa Postal 5008, 58051-970, Jo\~ao Pessoa, Para\'{\i}ba, Brazil}
\emailAdd{flavio2.bombacigno@uv.es}
\emailAdd{gonzalo.olmo@uv.es}
\emailAdd{orazi.emanuele@gmail.com}
\emailAdd{pporfirio@fisica.ufpb.br}

\abstract{We consider a metric-affine formulation of Chern-Simons modified gravity in $2+1$ dimensions. The theory is built requiring projective invariance, and the structure of the equations is analyzed using a decomposition in terms of scalar, vectorial, and purely tensorial objects. This approach allows us to implement a perturbative approach to study the corrections that emerge around a given background solution, for which we consider a BTZ charged, non-rotating metric. We show that conditions on model parameters are necessary to keep perturbations under control, yielding a rotating solution with a constant angular momentum and magnetic field at the horizon, and a smooth decay further away. We comment on the possibility of going beyond the leading order in perturbations and on its dynamical implications.   }
\makeatletter
\gdef\@fpheader{}
\makeatother

\begin{document}

\maketitle
\flushbottom

\section{Introduction}\label{sec: 1}

The exploration of alternative theories of gravity has flourished in recent years, boosted by new observational and theoretical discoveries. The accelerated expansion of the Universe \cite{SupernovaSearchTeam:1998fmf, SupernovaCosmologyProject:1998vns, SupernovaSearchTeam:2004lze}, the dynamics of galaxies in clusters \cite{Persic:1995ru}, and gravitational lensing phenomena \cite{Liebes:1964zz} can find an explanation within the framework of General Relativity (GR) only if two as-yet undetected matter-energy sources of unknown nature are assumed to dominate the gravitational dynamics at large scales. The lack of any direct observational signature of such \textit{dark} sources, together with the difficulties in disentangling their effects from potential modifications of GR at those scales, naturally justifies the exploration of gravity theories beyond GR. Also, in the strong-field regime, recent observations have opened a complementary avenue to search for new physics in the vicinity of compact objects, such as black holes. In particular, the detection of the supermassive compact object ({\it Saggitarius} A$^*$) at the center of the Milky Way \cite{EventHorizonTelescope:2022wkp, EventHorizonTelescope:2022apq, EventHorizonTelescope:2022wok}, together with precision measurements of stellar orbits and the growing catalogue of gravitational-wave events from compact-binary mergers, provide access to explore gravity regimes far beyond those available in weak-field tests. These strong-field observations can therefore be used to constrain (or reveal) departures from GR through signatures in spacetime geometry near horizons and the propagation of gravitational waves \cite{LIGOScientific:2016aoc, LIGOScientific:2016sjg, LIGOScientific:2016dsl, LIGOScientific:2017ync}. These large-scale/strong-field regime modifications of gravity are natural from a theoretical perspective if GR is regarded as an effective theory that requires some kind of completion in the ultraviolet and infrared sectors \cite{Donoghue:1994dn,Petrov:2020wgy}. For this same reason, high-energy extensions aimed at addressing questions related to the early Universe and black hole interiors have also received much attention.

Theoretical developments in the context of anomaly cancellations in quantum field theory \cite{Alvarez-Gaume:1983ihn} have also motivated extensions of GR able to generate parity violation \cite{Jackiw:2003pm, Alexander:2009tp}. Theories including topological contributions such as Chern--Simons terms in their actions have attracted much attention due to their generic appearance wherever the principles of gauge invariance are present \cite{Porfirio:2016nzr, Porfirio:2016ssx, Boudet:2022wmb, Olmo:2023tyi, Nascimento:2024dkt, Nascimento:2025qek}. These effective theories arise naturally in scenarios as different as particle physics \cite{Mariz:2004cv, Mariz:2007gf, Gomes:2008an}, loop quantum gravity \cite{Taveras:2008yf}, and string theory \cite{Polchinski:1998rr, Gukov:2003cy}, which exemplifies their relevance in the high-energy regime.

Although most extensions of General Relativity (GR) introduce new physical degrees of freedom to accommodate a wide range of unexplained phenomena, enlarging the field content generically increases the model-building freedom and may reduce the predictivity of the theory. An alternative, theoretically robust strategy is to retain the original propagating degrees of freedom while restricting the dynamics to second-order field equations, thereby avoiding the instabilities commonly associated with higher-order theories. In this regard, the Palatini (or metric–affine) formulation of gravity provides a particularly appealing framework. In this approach, the metric and the affine connection are treated as independent fields (see, e.g., the review \cite{Olmo:2011uz}). This independence enlarges the off-shell geometric structure while preserving the on-shell propagating degrees of freedom of GR. As a result, novel interaction terms emerge at the effective level \cite{Latorre:2017uve,Iosifidis:2020dck,Delhom:2019wir,Delhom:2020hkb,Iosifidis:2021bad}, even though no additional dynamical fields are introduced. The key point is that modifying the relation between metric and connection alters the gravitational dynamics without changing the underlying field content, possibly inducing nonlinear interactions. Moreover, imposing specific constraints on the affine connection naturally leads to torsion- or non-metricity-based formulations of gravity. Prominent examples include teleparallel gravity and symmetric teleparallel gravity, where curvature is replaced by torsion or non-metricity as the mediator of gravitational interactions. These frameworks, as well as their extensions, further illustrate how alternative geometric descriptions can produce nontrivial dynamics without increasing the number of degrees of freedom \cite{Maluf:2013gaa,BeltranJimenez:2017tkd,BeltranJimenez:2019odq,Heisenberg:2023lru}. Importantly, preserving the original degrees of freedom does not preclude the emergence of rich nonlinear phenomena. On the contrary, the modified couplings permitted in these approaches can generate significant deviations from GR in high-curvature regimes. At the same time, the restriction to second-order field equations prevents the appearance of Ostrogradsky-type instabilities, eliminating the need for additional screening mechanisms to suppress unwanted extra modes. Remarkably, such theories have demonstrated an unexpected capacity to resolve classical spacetime singularities. In particular, various models within the Palatini and related metric–affine frameworks have been shown to regularize both cosmological singularities and black hole interiors, replacing them with non-singular geometries under physically reasonable conditions \cite{Olmo:2008nf,Delhom:2023xxp,Maso-Ferrando:2023wtz,Maso-Ferrando:2023nju}. This singularity-resolution mechanism arises from the nonlinear structure of the modified gravitational dynamics rather than from the introduction of new propagating fields, highlighting the theoretical and phenomenological relevance of these approaches.

Interestingly, while the Pontryagin term of Chern--Simons theories is defined in terms of the connection of the gauge fields, its implementation in the gravitational scenario has experienced an unnatural departure from its motivating context. Rather than considering that term as genuinely based on the affine connection, as its construction suggests, a second-order approach has been systematically imposed by assuming that the connection is defined by the Christoffel symbols of the metric. This relegates the role of the connection to a merely notational appearance, putting all the physics on the metric side. The exploration of this type of theory from a metric-affine perspective could be relevant since, as it has been well established in recent years, metric and Palatini formulations of a given theory generically exhibit inequivalent dynamical equations. In the case of Chern–Simons gravity, the limited attempts to analyze its dynamics within the metric–affine framework have so far failed to provide a clarification of this issue beyond leading order in perturbation theory and typically rely on substantial simplifying assumptions \cite{Boudet:2022wmb,Sulantay:2022sag}. Moreover, as emphasized in \cite{Deser}, even in the $2+1$-dimensional case the resulting field equations are considerably more involved than in the standard metric formulation, thereby obstructing further progress along these lines.

{In general, gravitational theories in lower dimensions provide a valuable theoretical framework for exploring fundamental aspects of gravity in a simplified mathematical setting. Although it is a well-established result that GR in $2+1$ dimensions is a topological theory and it does not exhibit local propagating degrees of freedom \cite{Deser:1983tn}, the reduced dimensionality enables a detailed analysis of several nontrivial features that are often obscured in four dimensions, including canonical quantization, boundary dynamics, black hole solutions, and holography. Within this context, particular attention has been devoted to models that reintroduce propagating degrees of freedom. Prominent examples are displayed by Topologically Massive Gravity (TMG) \cite{Deser:1981wh} and its subsequent parity-preserving extension New Massive Gravity (NMG) \cite{Bergshoeff:2009hq}, where the Einstein–Hilbert action with a cosmological constant is supplemented, respectively, by a gravitational Lorentz–Chern–Simons term or specific curvature-squared contractions. However, the appearance in both models of negative central charges for the conformal field theory CFT$_2$, in the presence of positive-energy spin-2 modes in the dual AdS$_3$ bulk vacuum, led to the formulation of the Minimal Massive Gravity and its further extensions \cite{Hohm:2012vh,Bergshoeff:2014pca,Deger:2023eah}, in an attempt to deal with non-unitary problems at the boundary.}

In this paper, we reconsider the problem of Einstein gravity with a gravitational Chern--Simons term in $2+1$ dimensions using the Palatini formulation and find a way to deal with the connection equation, allowing us to solve it perturbatively. {We build our intuition on our previous analysis for the full $3+1$ case \cite{Boudet:2022wmb}, and we decide to treat perturbatively the resulting equations for the metric as well.} The zeroth-order equation reduces to the Einstein equation, while first-order corrections to the Einstein equation appear as unconventional terms involving contractions between the energy-momentum tensor and the Levi-Civita one. As an illustration of the effects of the new dynamics, we consider the gravitational field generated by a static electric charge. {We show how this configuration can be adopted as a seed solution for the perturbative corrections, which turn out to be consistent with a deformed charged Bañados-Teitelboim-Zanelli (BTZ) black-hole in the slow-rotating limit \cite{Banados:1992wn,Martinez:1999qi,Bueno:2021krl}. In particular, we show how rotational effects are explicitly sustained by the parity violating corrections generated by the Chern-Simons term, triggering in turn a non-trivial magnetic field.}

The paper is organized as follows. In \autoref{sec: 2}, we present a projectively invariant  metric-affine formulation of the Chern-Simons theory of gravity. In \autoref{sec: 3},  we derive the field equations. In \autoref{sec: 4}, we obtain a slowly rotating BTZ solution within the Chern-Simons theory of gravity. Finally, in \autoref{sec: 5}, we present our conclusions.

Notation is established as follows. The gravitational coupling is denoted by $\kappa^2 = 8 \pi G_3$, where $G_3$ is the $3$-dimensional Newton constant, which in natural units $\hbar = 1 = c$ has dimensions of a length; space-time signature is chosen to be mostly plus, i.e. $\eta_{\mu\nu} = (-1,1,1)$; the completely antisymmetric symbol $\epsilon^{\alpha\beta\gamma}$ is related to the Levi-Civita tensor $\varepsilon^{\alpha\beta\gamma}$ by $\epsilon^{\alpha\beta\gamma}=\sqrt{-g}\varepsilon^{\alpha\beta\gamma}$.

\section{The Palatini Chern-Simons theory of gravity}\label{sec: 2}
The starting point of our analysis is a Palatini formulation of the Chern-Simons model of gravity in $2+1$ dimensions, i.e.
\begin{equation}
    S=\act \leri{\pal-2\Lambda}+S_{CS}+S_M,
    \label{eq: action}
\end{equation}
where the Chern-Simons term is defined as
\begin{equation}
    S_{CS}=-\frac{1}{8\mu \kappa^2}\int d^3x\; \epsilon^{\alpha\beta\gamma}\leri{\pal\indices{^\rho_{\sigma\beta\gamma}}\Gamma\indices{^\sigma_{\rho\alpha}}-\frac{\Delta}{3}\hat{\pal}\indices{_{\beta\gamma}}\Gamma\indices{^\rho_{\rho\alpha}}+\frac{2}{3}\Gamma\indices{^\rho_{\sigma\alpha}}\Gamma\indices{^\sigma_{\tau\beta}}\Gamma\indices{^\tau_{\rho\gamma}}},
    \label{eq: CS term def}
\end{equation}
and it is controlled by the coupling $\mu$. $S_M$ is the action for the matter fields and $\Lambda$ the cosmological constant. As opposed to its purely metric formulation, here the connection must be understood as an independent geometric entity with respect to the metric field, and the Riemann tensor and the covariant derivative are formally defined, respectively, as:
\begin{equation}
\mathcal{R}\indices{^\rho_{\mu\sigma\nu}}=\partial_\sigma\Gamma\indices{^\rho_{\mu\nu}}-\partial_\nu\Gamma\indices{^\rho_{\mu\sigma}}+\Gamma\indices{^\rho_{\tau\sigma}}\Gamma\indices{^\tau_{\mu\nu}}-\Gamma\indices{^\rho_{\tau\nu}}\Gamma\indices{^\tau_{\mu\sigma}},
\end{equation}
and
\begin{equation}
    \nabla_\mu V\indices{^\rho_\sigma}=\partial_\mu V\indices{^\rho_\sigma}+\Gamma\indices{^\rho_{\lambda\mu}}V\indices{^\lambda_\sigma}-\Gamma\indices{^\lambda_{\sigma\mu}}V\indices{^\rho_\lambda}.
\end{equation}
We note that compared to the standard metric case, the Chern-Simons term in \eqref{eq: CS term def} exhibits an extra contribution led by the parameter $\Delta$ and containing the so-called homothetic curvature
\begin{equation}
    \hat{\pal}_{\mu\nu}\equiv \mathcal{R}\indices{^\alpha_{\alpha\mu\nu}},
\end{equation}
which represents the additional trace of the Riemann tensor solely existing in a metric-affine formalism (the other being the ordinary Ricci tensor). The inclusion of such a term is motivated by the requirement that the theory be invariant under projective transformations of the affine connection, under which the connection transforms according to the projective shift
\begin{equation}
\Gamma\indices{^\rho_{\mu\nu}}\rightarrow\tilde{\Gamma}\indices{^\rho_{\mu\nu}}=\Gamma\indices{^\rho_{\mu\nu}}+\delta\indices{^\rho_\mu}\xi_\nu,
    \label{projective}
\end{equation}
where $\xi_\mu$ is an a priori undefined one-form. Now, since for \eqref{projective} the Riemann tensor transforms as
\begin{equation}
\mathcal{R}\indices{^\rho_{\mu\sigma\nu}}\rightarrow\tilde{\mathcal{R}}\indices{^\rho_{\mu\sigma\nu}}=\mathcal{R}\indices{^\rho_{\mu\sigma\nu}}-\delta\indices{^\rho_\mu}\partial_\sigma\xi_\nu+\delta\indices{^\rho_\mu}\partial_\nu\xi_\sigma,
    \label{projective transformation Riemann}
\end{equation} 
it is easy to see that the Ricci scalar is invariant under \eqref{projective}, so that projective transformations can be considered, within the context of a Palatini formulation of General Relativity, an additional symmetry (see \cite{Hehl:1994ue,Sauro:2022hoh,Iosifidis:2019fsh} for details) that can be partially promoted to a gauge symmetry \cite{Olmo:2022ops}. However, in the presence of higher-order corrections in the action, as is the case of Chern-Simons gravity, projective invariance can be violated, leading to the appearance of dynamical instabilities \cite{BeltranJimenez2019,BeltranJimenez:2020sqf}. A projective invariant description of Chern-Simons gravity cannot be achieved therefore by simply promoting the original metric action to its metric-affine counterpart, and additional terms in curvature are in general expected. The homothetic curvature term is then purposely designed to restore projective invariance when $\Delta=1$, in that its variation under \eqref{projective} exactly cancels out the contribution arising from the Riemann tensor in \eqref{eq: CS term def}. We observe that an analogous term arises also when a projective invariant formulation of Chern-Simons gravity is pursued in $3+1$ dimensions (see \cite{Boudet:2022wmb}), where homothetic contributions can even preserve the topological behavior of the original Chern-Simons term in four dimensions.\\

We start our analysis by observing that the Chern-Simons term does not depend in principle on the metric, in that it is built on the completely anti-symmetric Levi-Civita contravariant symbol $\epsilon^{\rho\mu\nu}$, which upon variation with respect to the metric only contributes with a boundary term. This implies that by varying \eqref{eq: action} with respect to $g_{\mu\nu}$ one simply obtains
\begin{equation}
    \pal_{(\mu\nu)}-\frac{1}{2}g_{\mu\nu}\leri{\pal-2\Lambda}=\kappa^2 \tau_{\mu\nu},
    \label{eq: metric equation}
\end{equation}
with $\tau_{\mu\nu}$ denoting the energy momentum tensor of the matter fields in \eqref{eq: action}. By resorting to the decomposition of the affine connection in terms of its Levi-Civita component and the distorsion tensor (see definitions and notation in App.~\ref{app: A}), it is possible to single out in \eqref{eq: metric equation} the proper metric Einstein tensor from the non-Riemannian contribution, i.e.
\begin{equation}
G_{\mu\nu}+A_{\mu\nu}+g_{\mu\nu}\Lambda=\kappa^2\tau_{\mu\nu},
\label{eq: metric equation dec}
\end{equation}
where the expression for $A_{\mu\nu}$ is given in \eqref{eq: A tensor affine}. Its explicit form can be determined only once the equation for the connection has been solved, which by applying the techniques and the formalism introduced in App.~\ref{app: A}, and by assuming that matter does not depend on the connection\footnote{This choice amounts to consider a proper Palatini formulation, where the hypermomentum tensor is identically vanishing. For scenarios with non trivial hypermomentum see for example \cite{Iosifidis:2020gth,Iosifidis:2021nra}}, turns out to be \begin{align}
    &-\nabla_\lambda\leri{\sqrt{-g} g^{\mu\nu}}+\delta^\nu_\lambda\nabla_\rho\leri{\sqrt{-g} g^{\mu\rho}}+\sqrt{-g}\leri{g^{\mu\nu}T\indices{^\tau_{\lambda\tau}}-\delta\indices{^\nu_\lambda}T\indices{^{\tau\mu}_\tau}+T\indices{^{\nu\mu}_\lambda}}=\nonumber\\
    &=\frac{\epsilon^{\beta\gamma\nu}}{2\mu}\leri{\mathcal{R}\indices{^\mu_{\lambda\beta\gamma}}-\frac{\Delta}{3}\delta\indices{^\mu_\lambda}\hat{\mathcal{R}}\indices{_{\beta\gamma}}} .
    \label{eq: connection general}
\end{align}
Obtaining a solution of \eqref{eq: connection general} is, in general, a challenging task, due to the appearance of the full metric-affine Riemann tensor on the right-hand side of the equation, carrying quadratic and covariant derivatives terms of the connection components. However, relying solely on the behavior of the metric Riemann tensor in $2+1$ dimensions, some general properties can already be inferred, related to the nature of the corrections which are expected to arise at the effective level. It is known, indeed, that in $2+1$ dimensions the metric Weyl tensor is identically vanishing, so that the metric Riemann tensor can be completely rewritten in terms of the metric Ricci scalar and the Ricci tensor, i.e.
\begin{equation}
    R_{\mu\nu\rho\sigma}=g_{\nu\sigma}R_{\mu\rho}-g_{\nu\rho}R_{\mu\sigma}+g_{\mu\rho}R_{\nu\sigma}-g_{\mu\sigma}R_{\nu\rho}-\frac{R}{2}\leri{g_{\mu\rho}g_{\nu\sigma}-g_{\mu\sigma}g_{\nu\rho}}.
    \label{eq: metric riemann dec}
\end{equation}
These quantities can be in turn expressed in terms of the matter fields along with the tensor $A_{\mu\nu}$, as it can be appreciated by evaluating the trace of \eqref{eq: metric equation dec}, i.e.
\begin{equation}
    R=-2\kappa^2\tau+6\Lambda+2A,\quad R_{\mu\nu}=\kappa^2 (\tau_{\mu\nu}-g_{\mu\nu}\tau)+2g_{\mu\nu}\Lambda-A_{\mu\nu}+g_{\mu\nu}A.
    \label{eq: ricci matter}
\end{equation}
Thus, in order to deal with the complexity and the nonlinearity of the model, we choose to pursue the strategy of considering the Chern-Simons term as a small correction to the background theory, embodied by a Palatini formulation of General Relativity in $2+1$ dimensions. This perspective allows us to look for perturbative solutions ruled by the parameter $\epsilon \equiv \mu^{-1}$, with the idea of elucidating in the linearized theory the main contribution of the modified theoretical framework. Based on these assumptions, we see that even if the hypermomentum is set to zero from the very beginning, the connection still depends on the matter once the metric part of the Riemann tensor is rewritten in \eqref{eq: connection general} by means of \eqref{eq: metric riemann dec}. This indicates, therefore, that the original Einstein equations are expected to be supplemented with additional source terms, carrying contributions built out of the energy momentum tensor components. 

One should note that, in principle, modifications to the dynamical structure of the theory are also attainable, due to the differential nature of the equation for the connection, possibly involving the dependence of the connection on the metric field and its derivatives. However, as will be discussed in detail in \autoref{sec: 3}, torsion and non-metricity can never actually sustain additional degrees of freedom, in that the perturbative structure in $\epsilon$ of \eqref{eq: connection general} forces the components of the connection to be algebraically determined by the lower order solutions of the fields.

\section{The equation for the connection}\label{sec: 3}
The general equation displayed by \eqref{eq: connection general} can be rearranged in terms of the irreducible components of torsion and non-metricity (see definitions in App.~\ref{app: A}) as
\begin{align}
    &\frac{1}{2}Q_\lambda g_{\mu\nu}-Q_{\lambda\mu\nu}-g_{\nu\lambda}\leri{\frac{1}{2}Q_\mu-P_\mu}+g_{\mu\nu}T_\lambda-g_{\nu\lambda}T_\mu+T_{\nu\mu\lambda}=\frac{\epsilon}{2}\varepsilon^{\beta\gamma\nu}\leri{\mathcal{R}_{\mu\lambda\beta\gamma}-\frac{\Delta}{3}g_{\mu\lambda}\hat{\mathcal{R}}_{\beta\gamma}},
    \label{eq: connection general component}
\end{align}
where we lowered all the indices and used the identity $\nabla_\mu\sqrt{-g}=-\frac{\sqrt{-g}}{2}Q_\mu$. By successive contractions with $g^{\mu\lambda},\;g^{\nu\lambda},\;g^{\mu\nu}$ and $\varepsilon^{\nu\mu\lambda}$, it is possible to extract from \eqref{eq: connection general component} four equations for the three vector and the pseudoscalar parts of torsion and nonmetricity, i.e.
\begin{align}
    &\frac{\epsilon}{2}\varepsilon^{\beta\gamma\nu}(1-\Delta)\hat{\mathcal{R}}_{\beta\gamma}=0
    \label{eq: contraction ml}\\
    &2P_\mu-Q_\mu-T_\mu=\frac{\epsilon}{2}\varepsilon^{\alpha\beta\gamma}\leri{\pal_{\mu\alpha\beta\gamma}-\frac{\Delta}{3}g_{\mu\alpha}\hat{\pal}_{\beta\gamma}}
    \label{eq: contraction nl}\\&P_\mu+T_\mu=\frac{\epsilon}{2}\varepsilon^{\alpha\beta\gamma}\leri{\pal_{\alpha\mu\beta\gamma}-\frac{\Delta}{3}g_{\alpha\mu}\hat{\pal}_{\beta\gamma}}
\label{eq: contraction mn}\\
&\Theta=-\epsilon\pal.
    \label{eq: contraction eps}
\end{align}
We stress the fact that by virtue of \eqref{eq: metric equation}, the pseudoscalar part of torsion is completely determined in terms of the matter, i.e.
\begin{equation}
   \Theta=\epsilon\leri{2\kappa^2\tau-6\Lambda},
\end{equation}
a result holding exactly, irrespectively of the perturbative orders, which only concur in determining the structure of the energy momentum trace. Concerning instead the vector part of the connection, we note that under a projective transformation in $2+1$ dimensions the traces of torsion and nonmetricity transform as
\begin{align}
    &T^\rho\rightarrow\tilde{T}^\rho= T^\rho-2\xi^{\rho},\\
    &Q^\rho\rightarrow\tilde{Q}^\rho= Q^\rho+6\xi^{\rho},\\
    &P^\rho\rightarrow\tilde{P}^\rho= P^\rho+2\xi^{\rho},
    \label{eq: vector proj}
\end{align}
so that the l.h.s. of \eqref{eq: contraction nl} and \eqref{eq: contraction mn} are accordingly left invariant. Conversely, from \eqref{eq: contraction ml} we see that projective invariance is fully recovered only for $\Delta=1$, while for $\Delta=0$, formally corresponding to the original metric formulation, it is explicitly violated. In the following, we will exploit such a gauge freedom to set $T_\mu=0$ identically, to simplify the analysis and to retain only the non-metricity vectors.
\\By the inspection of \eqref{eq: contraction nl}, \eqref{eq: contraction mn} and \eqref{eq: contraction eps}, we immediately conclude that at the lowest order non-metricity traces and the pseudoscalar torsion are vanishing, leaving us with the equation for the tensor part
\begin{equation}
    q_{\nu\mu\lambda}-\Omega_{\lambda\mu\nu}=0.
\end{equation}
It is easy to see that this equation automatically implies $q_{\nu\mu\lambda}=\Omega_{\lambda\mu\nu}=0$ as well, as it can be appreciated by taking the symmetric part in $(\lambda,\mu)$, leading to $\Omega_{(\lambda\mu)\nu}=0$, which combined with the symmetry of the tensor $\Omega_{\rho\mu\nu}$ in the last two indices results in $\Omega_{\rho\mu\nu}=0$, and then $q_{\nu\mu\lambda}=0$. At the lowest order, therefore, the theory reduces to metric General Relativity in the presence of matter, i.e.
\begin{equation}
G_{\mu\nu}+g_{\mu\nu}\Lambda=\kappa^2 \tau_{\mu\nu},
    \label{eq: metric bg}
\end{equation}
in agreement with our initial hypothesis. Moving to the first order in the perturbative expansion, the system of equations \eqref{eq: contraction mn}-\eqref{eq: contraction nl} instead yields the following solutions 
\begin{align}
    \delta Q_\mu=3\delta P_\mu=-\frac{3\epsilon}{2}\varepsilon^{\alpha\beta\gamma}R_{\mu\alpha\beta\gamma},
\end{align}
which, taking into account \eqref{eq: metric riemann dec} and \eqref{eq: ricci matter}, it is immediate to see to vanish, i.e. $\delta Q_\mu = \delta P_\mu = 0$. We note that such a result can be directly inferred also resorting to the Bianchi identity for the metric Riemann tensor. Hence, plugging back these results into the original equation for the connection, we end up with the following relation for the tensor part of torsion and non-metricity
\begin{equation}
    \delta T_{\nu\mu\lambda} - \delta \Omega_{\lambda\mu\nu}=\frac{\epsilon}{2}\varepsilon\indices{^{\alpha\beta}_\nu}R_{\mu\lambda\alpha\beta}.
\end{equation}
Again, by taking the symmetric part in the $(\lambda,\mu)$ indices, it is easy to show that $\delta\Omega_{\lambda\mu\nu}=0$, leaving us with a clear expression for the torsion tensor, i.e.
\begin{equation}
    \delta T_{\nu\mu\lambda}=\frac{\epsilon}{2}\varepsilon\indices{^{\alpha\beta}_\nu}R_{\mu\lambda\alpha\beta},
\end{equation}
from which the purely tensor part can be extracted as $\delta q_{\nu\mu\lambda}=\delta T_{\nu\mu\lambda}+\frac{\delta \Theta}{6}\varepsilon_{\nu\mu\lambda}$. Finally, substituting the expressions for the Ricci tensor and the Ricci scalar in terms of the energy momentum tensor, it is possible to entirely rewrite the torsion perturbation as a function of matter, i.e.
\begin{equation}
    \delta T_{\rho\mu\nu}= \frac{\kappa^2\epsilon}{2}\leri{\varepsilon\indices{^\lambda_{\rho\mu}}\tau_{\nu\lambda}-\varepsilon\indices{^\lambda_{\rho\nu}}\tau_{\mu\lambda}-\varepsilon_{\rho\mu\nu}\leri{\tau-\frac{\Lambda}{\kappa^2}}}.
    \label{eq: solution torsion matter}
\end{equation}
As briefly mentioned at the end of \autoref{sec: 2}, a similar result implies that when we look at the next perturbative orders, the connection components appearing on the l.h.s. of \eqref{eq: connection general component} are algebraically determined only by the lower order quantities on the r.h.s., by virtue of the overall factor $\epsilon$ ruling the Chern-Simons term. This property, together with the explicit solution displayed in \eqref{eq: solution torsion matter}, ensures that torsion and non-metricity can be iteratively constructed from the background configuration for the metric field and the energy momentum tensor, so that a truly differential equation for the connection never shows up and additional degrees of freedom cannot be excited (see \cite{Baldazzi:2021kaf,Mikura:2023ruz,Mikura:2024mji,Percacci:2025oxw}). In turn, this also implies that at each order the equation for the corresponding metric perturbation is only affected in the source part, in that no higher-derivative operators can be introduced by the affine structure. Indeed, the covariant derivatives contained in the tensor $A_{\mu\nu}$ can only act on the metric field and the energy momentum tensor of the previous orders, so that they can be globally considered as modified source terms. This is in contrast with the original purely metric formulation in TMG \cite{Adami:2021sko}, where the inclusion of a Chern-Simons term introduces a new mass scale and results in a (linearized) third-order wave-equation, which propagates a single massive mode of helicity $\pm 2$, the sign depending on the sign of the Chern-Simons term.\\

We now restrict our focus to the linearized metric field equation, i.e.
\begin{equation}
    \delta G_{\mu\nu}+\delta g_{\mu\nu}\Lambda+\delta A_{\mu\nu}=\kappa^2 \delta \tau_{\mu\nu},
\end{equation}
where $\delta A_{\mu\nu}$ is displayed by
\begin{equation}
    \delta A_{\mu\nu}=\Bar{D}\indices{_\rho} \delta N\indices{^\rho_{(\mu\nu)}}-\Bar{D}\indices{_{(\nu}}\delta N\indices{^\rho_{\mu)\rho}}-\frac{1}{2}g_{\mu\nu}\Bar{D}\indices{_{\rho}}\leri{\delta N\indices{^{\rho\sigma}_\sigma}-\delta N\indices{^{\sigma\rho}_{\sigma}}},
\end{equation}
with a bar over the metric covariant derivative denoting evaluation on the background Levi-Civita connection. In particular, since in our case non-metricity is completely vanishing and the trace of torsion is set to zero for projective symmetry, it can be checked that the tensor perturbation $\delta A_{\mu\nu}$ takes the simple form
\begin{equation}
    \delta A_{\mu\nu}=\Bar{D}_\rho \delta T\indices{_{(\mu\nu)}^\rho},
\end{equation}
so that the equation for the metric perturbation can be eventually rearranged as
\begin{equation}
    \delta G_{\mu\nu}+\delta g_{\mu\nu}\Lambda=\kappa^2\leri{\delta \tau_{\mu\nu}+\epsilon\,\varepsilon\indices{^{\rho\sigma}_{(\mu}}\Bar{D}\indices{_\rho} \tau\indices{_{\nu)\sigma}}}.
    \label{eq: metric linearized mod source}
\end{equation}
We see therefore that at the first perturbative level, the global effect of the Chern-Simons modification amounts to an additional source term, carrying derivatives of the background matter distribution. Note that on dimensional grounds, $\epsilon$ represents a length scale. 

\section{Slow rotating BTZ charged solution}\label{sec: 4}
In the following, we consider the background configuration generated by \eqref{eq: action}, when the matter Lagrangian is presented by the electromagnetic field $A_\mu(x)$, i.e.
\begin{equation}
    S=\act \leri{\pal-2\Lambda-F_{\mu\nu}F^{\mu\nu}}+S_{CS},
    \label{eq: action charged}
\end{equation}
with $F_{\mu\nu}:=\partial_\mu A_\nu-\partial_\nu A_\mu$. In this scenario,  a physically relevant background  metric is given by the well-known charged BTZ black hole solution \cite{Banados:1992wn}, characterized by the line element
\begin{equation}
    ds^2=-f(r)dt^2+\frac{1}{f(r)}dr^2+r^2d\phi^2,
\end{equation}
with\footnote{On dimensional grounds, here $M$ and $\kappa^2Q_1^2$ are dimensionless, $\Lambda$ has dimensions of inverse squared length, and $r_0$ is an arbitrary length scale that makes dimensionless the argument of the logarithm.  }
\begin{equation}
    f(r)=-M-\Lambda r^2-2\kappa^2Q_1^2\ln r/r_0,
\end{equation}
where $\Lambda<0$ and satisfying the Einstein equations
\begin{align}
&G_{\mu\nu}+g_{\mu\nu}\Lambda=\kappa^2 \tau_{\mu\nu}.
\end{align}
The energy-momentum tensor $\tau_{\mu\nu}$ for the electromagnetic field is given by
\begin{equation}
   \tau_{\mu\nu}=2F_{\mu\alpha}{F_\nu}^\alpha-\frac{1}{2}g_{\mu\nu}F,
\end{equation}
where gauge invariance allows us to rewrite the electromagnetic tensor as
\begin{equation}
    F_{\mu\nu}=E(r)\leri{\delta\indices{^r_\mu}\delta\indices{^t_\nu}-\delta\indices{^r_\nu}\delta\indices{^t_\mu}}=\frac{Q_1}{r}\leri{\delta\indices{^r_\mu}\delta\indices{^t_\nu}-\delta\indices{^r_\nu}\delta\indices{^t_\mu}},
\end{equation}
with $E(r)=Q_1/r$ representing the radial electric field with charge $Q_1$. 
In order to elucidate the nature of the correction induced by the Chern-Simons terms, we make the hypothesis that at the first order in perturbation the metric exhibits an off-diagonal term related to a slow-rotation of the spacetime. We show this rotation to be sustained by a magnetic field correction in the electromagnetic tensor, in turn triggering a non diagonal energy momentum tensor. In particular, we clarify how the modified source term exactly retains the same off-diagonal structure, guaranteeing a genuine departure from the standard general relativity case in the presence of rotation. Namely, we consider the perturbed line element\footnote{Here $\epsilon$ has dimensions of length, $\phi$ is dimensionless, and $\omega(r)$ has dimensions of inverse squared length.}
\begin{align}
    ds^2=-f(r)dt^2+\frac{1}{f(r)}dr^2+r^2d\phi^2{\color{red}+2\epsilon r^2\omega(r)dtd\phi},
\end{align}
and the perturbed electromagnetic tensor
\begin{equation}
    F_{\mu\nu}=E(r)\leri{\delta\indices{^r_\mu}\delta\indices{^t_\nu}-\delta\indices{^r_\nu}\delta\indices{^t_\mu}}{\color{red}+\epsilon B(r)\leri{\delta\indices{^r_\mu}\delta\indices{^\phi_\nu}-\delta\indices{^r_\nu}\delta\indices{^\phi_\mu}}}.
\end{equation}
For this configuration, the only Einstein tensor component receiving a linear correction in $\epsilon$ is the element $t\phi$, i.e.
\begin{align}
    &\Bar{G}_{tt}=-\frac{f(r)f'(r)}{2r}+\mathcal{O}(\epsilon^2)\\
    &\Bar{G}_{rr}=\frac{f'(r)}{2rf(r)}+\mathcal{O}(\epsilon^2)\\
    &\Bar{G}_{\phi\phi}=\frac{r^2 f''(r)}{2}+\mathcal{O}(\epsilon^2)\\
    &{\color{red}\Bar{G}_{t\phi}=-\frac{\epsilon}{2}\leri{\omega''(r)+\frac{3}{r}\omega'(r)-\frac{f''(r)}{f(r)}\omega(r)}f(r)r^2+\mathcal{O}(\epsilon^2)}
\end{align}
in agreement with the perturbed energy momentum tensor components
\begin{align}
    &\kappa^2\tau_{tt}=f(r)E^2(r)+\mathcal{O}(\epsilon^2)\\
    &\kappa^2\tau_{rr}=-\frac{E^2(r)}{f(r)}+\mathcal{O}(\epsilon^2)\\
    &\kappa^2\tau_{\phi\phi}=r^2 E^2(r)+\mathcal{O}(\epsilon^2)\\
    &\kappa^2\tau_{tr}=\mathcal{O}(\epsilon^2)\\
    &\kappa^2\tau_{r\phi}=\mathcal{O}(\epsilon^3)\\
    &{\color{red} \kappa^2\tau_{t\phi}=\epsilon E^2(r)\leri{\frac{\omega(r)}{f(r)}+\frac{2}{r^2}\frac{B(r)}{E(r)}}f(r)r^2+\mathcal{O}(\epsilon^3)}.
\end{align}
We note that the electromagnetic invariant $F$ is left unchanged at order $\epsilon$, i.e. $F=F_{\mu\nu}F^{\mu\nu}=-2E^2(r)+\mathcal{O}(\epsilon^2)$. The Chern-Simons modification results instead in
\begin{equation}
\epsilon\kappa^2\varepsilon\indices{^{\rho\sigma}_{(t|}}\Bar{\nabla}\indices{_\rho} \tau\indices{_{|\phi)\sigma}}=\epsilon f(r)\frac{d}{dr}(rE^2(r))+\mathcal{O}(\epsilon^2),
\end{equation}
with all the other components of order $\epsilon^2$ or higher. From the Maxwell equation
\begin{equation}
    \partial_\mu\leri{\sqrt{-g}F^{\mu\nu}}=0,
\end{equation}
by selecting the component $\nu=\phi$ we obtain the equation for the magnetic field, i.e.
\begin{equation}
    \partial_r\leri{rg^{rr}g^{\phi t}F_{rt}+rg^{rr}g^{\phi\phi}F_{r\phi}}=\epsilon\partial_r\leri{r\omega(r)E(r)+\frac{f(r)}{r}B(r)}=0,
\end{equation}
which allows us to solve algebraically for $B(r)$, that is
\begin{equation}\label{eq:B}
    B(r)=\frac{r}{f(r)}\leri{Q_2-r\omega(r)E(r)}
\end{equation}
where $Q_2$ is an integration constant (with the same dimensions as $\omega$, which is an inverse squared length). Putting all together, we finally get from the linearized equation
\begin{equation}
    \delta G_{t\phi}+\delta g_{t\phi}\Lambda=\kappa^2\leri{\delta \tau_{t\phi}+\epsilon\varepsilon\indices{^{\rho\sigma}_{(t}}\Bar{\nabla}_\rho \tau_{\phi)\sigma}}
\end{equation}
the following differential equation for the metric function $\omega(r)$:
\begin{equation}\label{eq:wpert}
r^2f(r)\left(\omega''(r)+\frac{3}{r}\omega'(r)\right)-\leri{r^2f''(r)+2\kappa^2Q_1^2+2\Lambda r^2}\omega(r)+4\kappa^2 Q_1Q_2-\frac{2\kappa^2\delta f(r)Q_1^2}{r^2}=0,
\end{equation}
where $\delta=0,1$ is a book-keeping parameter which is zero for General Relativity and $1$ for our Chern-Simons model. Next we will explore the behavior of this equation in the two relevant asymptotic limits, namely $r\to \infty$ and $r\to r_H$, with $r_H$ representing the (outer) black hole horizon, which is determined by the largest root $r=r_H$ satisfying $f(r_H)=0$.  

\subsection{Far limit}

In the asymptotic region $r\to+\infty$, we can look for approximate analytic solutions of the equation
\begin{equation}
    \Lambda r^4 \omega''(r)+3\Lambda r^3 \omega'(r)+4\kappa^2 Q_1^2 \omega(r)=2\kappa^2Q_1(\delta\Lambda Q_1+2 Q_2) \ .
\end{equation}
To do so, it is convenient to perform a change of variable of the form $r=1/x$ such that the equation turns into 
\begin{equation}
    \ddot{\omega}-\frac{\dot{\omega}}{x}(x)+\frac{4\kappa^2 Q_1^2}{\Lambda} \omega=2\kappa^2Q_1(\delta\Lambda Q_1+2 Q_2) \ ,
\end{equation}
where $\dot{\omega}\equiv d\omega/dx$. It is now evident that this equation has a regular singular point around $x\to 0$ and the solution can be found by means of a Frobenius expansion. Before finding the general solution, we note that there is a constant term on the right-hand side of this equation that would imply a constant $\omega(r)$ as particular solution. However, such a term would lead to a quadratic growth for $g_{t\phi}$, indicating that this is not a perturbative solution. We must thus first impose the vanishing of that term, which establishes a constraint on the value of the integration constant $Q_2$. Thus, from now on we will set $Q_2=-\delta \Lambda Q_1/2$, which implies that the magnetic field is completely determined by the electric charge of the background field. The resulting solution for the homogeneous equation, takes the form 
\begin{equation}
    \omega(x)=c_1 x J_1\left(\frac{2 Q_1 x \kappa }{\sqrt{\Lambda }}\right)+c_2 x Y_1\left(\frac{2 Q_1 x \kappa }{\sqrt{\Lambda }}\right) \ ,
\end{equation}
where $J_1$ and $Y_1$ are the Bessel functions of the first and second kind, respectively. Expanding these functions around $x\to 0$, one finds that $x Y_1$ can cause quadratic and logarithmic divergences in the metric. To avoid this, the constant $c_2$ must be set to zero to have a well-behaved and decaying metric component $g_{t\phi}$. The solution in the $r\to \infty$ (or $x\to 0$) limit is thus given by $\omega(x)=c_1 x J_1\left(\frac{2 Q_1 x \kappa }{\sqrt{\Lambda }}\right)$, where $c_1$ is a free constant. With this result, we also find that the magnetic field in the far region behaves as $B(r)=\frac{Q_2}{\Lambda r} +Q_1 c_1 O(r^{-3})$. For completeness, we note that $x J_1\left(\frac{2 Q_1 x \kappa }{\sqrt{\Lambda }}\right)\approx \frac{\kappa  Q_1 x^2}{\sqrt{\Lambda }}+O(x^4)$. Since we are dealing with a negative $\Lambda=-1/l^2$, the arbitrary constant $c_1$ must be a purely imaginary number in order to produce a real solution for the metric perturbation, i.e., $c_1=i\tilde{c}_1$. A final comment regarding the asymptotic form of the $g_{t\phi}$ component is now in order. The functional form found is compatible with a constant angular momentum given by $\epsilon \ \omega(r)=-J_F/r^2$, where $J_F\equiv \epsilon \kappa Q_1 l \tilde{c}_1$.\\


\subsection{Near horizon limit}

We now assume configurations with at least one event horizon and consider an expansion of Eq.(\ref{eq:wpert}) around the outer horizon. We denote as $r_H$ the radial location of that horizon, which satisfies the equation $f(r_H)=0$ (see \autoref{app: B}). Considering locations parametrized as $r=r_H(1+\eta)$, where $\eta$ will play the role of the (dimensionless) radial displacement, (\ref{eq:wpert}) can be approximated as 
\begin{equation}
\omega_{\eta\eta}+3\omega_{\eta}+\frac{2\kappa^2Q_1^2}{\lambda_0 \eta}\omega=\frac{2\kappa^2Q_1 Q_2}{\lambda_0 \eta} \ ,
\end{equation}
where $\lambda_0\equiv 2r_H^4(\Lambda+\kappa^2Q_1^2/r_H^2)$. It is evident that this equation has a regular singular point at $\eta\to 0$ and a solution in terms of a Frobenius power series expansion is also possible. The result can be written as $\omega(\eta)=Q_2/Q_1+a_1 \eta +O(\eta^2)$, where $a_1$ is an arbitrary constant. Using the constraint on $Q_2$ that follows from the far field expansion, we simply get that $\omega(\eta)=-\delta \Lambda/2+a_1 \eta+O(\eta^2)$. This result shows that the expansion around the horizon is also finite and small, being consistent with the perturbative description. The resulting angular momentum on the horizon takes the form $J_H=-\epsilon \delta r_H^2/l^2$. The magnetic field requires special attention because, according to Eq.(\ref{eq:B}), it has a potential  divergence stemming from its denominator. However, taking into account the form of $\omega(\eta)$ found in this limit and the constraint on $Q_2$ coming from the far field expansion, a cancelation occurs in the numerator and the final result becomes finite, yielding 
\begin{equation}
    B(r_H)=\frac{r_H^2 Q_1 a_1}{2(\kappa^2Q_1^2+\Lambda r_H^2)} \ .
\end{equation}
This should be regarded as a highly nontrivial result that supports the consistency of our approach and choice of constraints among integration constants to get a well-behaved perturbative expansion. 

To conclude this section, we note that the relation between the undefined constants $a_1$ and $\tilde{c}_1$ could be determined numerically via the implementation of a shooting method for specific values of model parameters. A dedicated study of that type will be carried out elsewhere. \\

\section{Conclusion}\label{sec: 5}
In this work we have explored the structure of the equations of a Palatini formulation of Chern-Simons modified gravity in 2+1 dimensions. This theory is of interest on the grounds of high-energy modifications to Einstein's gravity, though the complexity of the resulting equations had precluded any progress in this direction so far. Thanks to the implementation of a decomposition of the torsion and non-metricity tensors in terms of scalar, vectorial, and purely tensorial components carried out by some of us in a 3+1 scenario in previous works, we have been able to tackle the problem from a perturbative perspective. Nonetheless, the peculiarities of the 2+1 scenario, with a vanishing Weyl tensor, allowed us to have a look beyond the perturbative level and realize that neither the torsion nor the non-metricity tensors generate new dynamical degrees of freedom. In fact, we saw that by an iterative approach one can progress to higher perturbative orders starting from a given background solution of GR. That formal aspect of the theory is, however, useless if the leading order perturbations are not well-behaved and, for that reason, we decided to explore the problem corresponding to a non-rotating, charged BTZ background solution. In this analysis, we found that a constraint between the charge of the background electric field and of an integration constant associated with the magnetic field (which emerges as a first-order correction) must be imposed in the far region to have the amplitude of perturbations under control in that limit. Surprisingly, it is precisely this constraint that guaranties a finite value of the magnetic field at the event horizon, which we interpret as a consistency check of our approach.
The resulting BTZ-corrected solution exhibits a finite angular momentum at the horizon which decays as $1/r^2$ at large distances, despite the lack of rotation of the initial background solution. 

The non-perturbative structure of the theory is not completely understood yet, though the results obtained here strongly motivate us to look for an iterative approach able to yield solutions of the complete problem. {Furthermore, the suspected non-perturbative reduced-order equations of motion with respect to the original metric formulation suggest that an independent connection may influence the relation between the AdS$_3$ bulk structure and the corresponding dual conformal field theory CFT$_2$. This feature can in principle directly impact the scaling dimensions of dual CFT$_2$ operators, offering a new perspective on the tension between the central charges on the boundary and the bulk modes as it typically arise in TMG and NMG.} We hope to discuss all these aspects in an upcoming publication. 

\acknowledgments
This work is supported by the Spanish National Grant PID2023-149560NB-C21, and the Severo Ochoa Excellence Grant CEX2023-001292-S, funded by MICIU/AEI/10.13039/501100011033 (“ERDF A way of making Europe”, “PGC Generacion de Conocimiento”) and FEDER, UE. The authors also acknowledge financial support from the project i-COOPB23096 (funded by CSIC). The paper is based upon work from COST Actions CosmoVerse CA21136 and CaLISTA CA21109, supported by COST (European Cooperation in Science and Technology).

\appendix
\section{A compendium of the metric-affine formalism in $2+1$ dimensions}\label{app: A}
For a generic non-Riemannian geometry, it is possible to introduce the notion of the torsion and nonmetricity tensors, which according our convention read respectively:
\begin{equation}
    \begin{split}
&T\indices{^\rho_{\mu\nu}}\equiv\Gamma\indices{^\rho_{\mu\nu}}-\Gamma\indices{^\rho_{\nu\mu}},\\
&Q\indices{_{\rho\mu\nu}}\equiv-\nabla_\rho g_{\mu\nu}.
    \end{split}
\end{equation}
In evaluating the equation of motion for the connection from \eqref{eq: action}, we use the generalized Palatini identity
\begin{equation}
\delta\mathcal{R}\indices{^\rho_{\mu\sigma\nu}}=\nabla_\sigma\delta\Gamma\indices{^\rho_{\mu\nu}}-\nabla_\nu\delta\Gamma\indices{^\rho_{\mu\sigma}}-T\indices{^\lambda_{\sigma\nu}}\delta\Gamma\indices{^\rho_{\mu\lambda}},
\end{equation}
and the property for vector densities
\begin{equation}
    \int d^3 x\; \nabla_\mu\leri{\sqrt{-g} V^\mu}=\int d^3x\; \partial_\mu\leri{\sqrt{-g}V^\mu}+\int d^3x\; \sqrt{-g}\;T\indices{^\rho_{\mu\rho}} V^\mu=\int d^3x\; \sqrt{-g}\;T\indices{^\rho_{\mu\rho}} V^\mu.
\end{equation}
We also rewrite torsion and nonmetricity in their irreducible parts (see \cite{Sauro:2022hoh} or the appendix of \cite{Boudet:2022wmb} for a comparison with the four dimensional case):
\begin{align}
    &T_{\mu\nu\rho} = \dfrac{1}{2}\left(T_{\nu}g_{\mu\rho}-T_{\rho}g_{\mu\nu}\right) -\dfrac{\Theta}{6} \varepsilon_{\mu\nu\rho} + q_{\mu\nu\rho},\label{torsion decomposition}\\
    &Q_{\rho\mu\nu}=\frac{2Q_\rho-P_\rho}{5}g_{\mu\nu}-\frac{Q_{(\mu}g_{\nu)\rho}-3P_{(\mu}g_{\nu)\rho}}{5}+\Omega_{\rho\mu\nu}.
    \label{non metricity decomposition}\\
    &N_{\rho\mu\nu}=\frac{1}{2}\leri{T_{\mu}g_{\nu\rho}-T_{\rho}g_{\mu\nu}-\dfrac{\Theta}{6} \varepsilon_{\rho\mu\nu} + q_{\rho\mu\nu}-q_{\mu\rho\nu}-q_{\nu\rho\mu} }+\nonumber\\
    &\quad\quad\quad+\frac{1}{2}\leri{-\frac{3Q_\rho-4P_\rho}{5}g_{\mu\nu}+\frac{4Q_{(\mu}g_{\nu)\rho}-2P_{(\mu}g_{\nu)\rho}}{5}+\Omega_{\mu\nu\rho}+\Omega_{\nu\mu\rho}-\Omega_{\rho\mu\nu}}
    \label{eq: distorsion decomposition}
\end{align}
We introduced the trace vectors
\begin{equation}
    T_{\mu} \equiv T \indices{^{\nu}_{\mu\nu}},\;Q_\rho=Q\indices{_\rho^\mu_\mu},\;P_\rho=Q\indices{^\mu_{\mu\rho}},
\end{equation}
the "axial" scalar trace for torsion
\begin{equation}
\Theta \equiv \varepsilon_{\mu\nu\rho}T^{\mu\nu\rho},
\end{equation}
and the traceless tensor parts $q_{\rho\mu\nu},\;\Omega_{\rho\mu\nu}$. The latter satisfy the additional properties $q_{\rho\mu\nu}=-q_{\rho\nu\mu},\;\epsilon^{\rho\mu\nu}q_{\rho\mu\nu}=0,\;\Omega_{\rho\mu\nu}=\Omega_{\rho\nu\mu}$. 
It is then possible to rewrite the affine connection as
\begin{equation}
    \Gamma\indices{^\rho_{\mu\nu}}=L\indices{^\rho_{\mu\nu}}+N\indices{^\rho_{\mu\nu}}=L\indices{^\rho_{\mu\nu}}+K\indices{^\rho_{\mu\nu}}+D\indices{^\rho_{\mu\nu}},
    \label{christoffel contorsion disformal}
\end{equation}
where $L\indices{^\rho_{\mu\nu}}$ denotes the Christoffel symbols, while the contorsion and disformal tensors are defined as
\begin{align}
    &K\indices{^\rho_{\mu\nu}}=\frac{1}{2}\leri{T\indices{^\rho_{\mu\nu}}-T\indices{_\mu^\rho_\nu}-T\indices{_\nu^\rho_\mu}}=-K\indices{_\mu^\rho_{\nu}},
    \label{decomposition contorsion}\\
    &D\indices{^\rho_{\mu\nu}}=\frac{1}{2}\leri{Q\indices{_{\mu\nu}^\rho}+Q\indices{_{\nu\mu}^\rho}-Q\indices{^\rho_{\mu\nu}}}=D\indices{^\rho_{\nu\mu}}.
    \label{decomposition disformal}
\end{align}
For a generic metric-affine structure the Riemann tensor is skew-symmetric only in its last two indices, so that we can take the contractions
\begin{align}
    \mathcal{R}_{\mu\nu}&\equiv\mathcal{R}\indices{^\alpha_{\mu\alpha\nu}},\\
    \hat{\mathcal{R}}_{\mu\nu}&\equiv \mathcal{R}\indices{^\alpha_{\alpha\mu\nu}},\\
    \mathcal{R}^\dag_{\mu\nu}&\equiv g_{\mu\tau}g^{\rho\sigma}\mathcal{R}\indices{^\tau_{\rho\sigma\nu}},
\end{align}
where the first one is the usual Ricci tensor, and the second one is called homothetic curvature. In particular, it is possible to show that the homothetic curvature rewrites as $\hat{\pal}_{\mu\nu}=\partial_{[\mu}Q_{\nu]}$. Eventually, the Riemann tensor can be expanded in terms of the distorsion tensor as
\begin{equation}
        \mathcal{R}_{\mu\rho\nu\sigma}=R_{\mu\rho\nu\sigma}+D{_{\nu}}N_{\mu\rho\sigma}-D{_{\sigma}} N_{\mu\rho\nu}+N_{\mu\lambda\nu}N\indices{^\lambda_{\rho\sigma}}-N_{\mu\lambda\sigma}N\indices{^\lambda_{\rho\nu}},
    \label{eq: decomposition riemann}
\end{equation}
where $R\indices{^\mu_{\nu\rho\sigma}}$ and $D{_{\mu}}$ are the metric quantities built from the Levi Civita connection. The affine Einstein tensor $\mathcal{G}_{\mu\nu}$ can be therefore rewritten as
\begin{equation}
    \mathcal{G}_{\mu\nu}={G}{_{\mu\nu}}+A_{\mu\nu},
\end{equation}
where we introduced the tensor \begin{equation}
\begin{split}
    A_{\mu\nu}\equiv\;& D\indices{_{\rho}}N\indices{^\rho_{(\mu\nu)}}-D\indices{_{(\nu}}N\indices{^\rho_{\mu)\rho}}+N\indices{^\rho_{\lambda\rho}}N\indices{^\lambda_{(\mu\nu)}}-N\indices{^\rho_{\lambda(\nu}}N\indices{^\lambda_{\mu)\rho}}+\\
    &-\frac{1}{2}g_{\mu\nu}\leri{D\indices{_{\rho}}\leri{N\indices{^{\rho\sigma}_\sigma}-N\indices{^{\sigma\rho}_{\sigma}}}+N\indices{^\rho_{\lambda\rho}}N\indices{^{\lambda\sigma}_\sigma}-N\indices{^\rho_{\lambda\sigma}}N\indices{^{\lambda\sigma}_\rho}},
\end{split}
\label{eq: A tensor affine}
\end{equation}
whose trace is given by
\begin{equation}
    A=-\frac{1}{2}\leri{D\indices{_{\rho}}\leri{N\indices{^{\rho\sigma}_\sigma}-N\indices{^{\sigma\rho}_{\sigma}}}+N\indices{^\rho_{\lambda\rho}}N\indices{^{\lambda\sigma}_\sigma}-N\indices{^\rho_{\lambda\sigma}}N\indices{^{\lambda\sigma}_\rho}}
\end{equation}

\section{Location of the inner and outer horizons}\label{app: B}
The position of the horizons is determined by solving $f(r_H)=0$, which by the change of variable
\begin{equation}
    y=\frac{\Lambda r_H^2}{\kappa^2 Q_1^2},
\end{equation}
with $y\in \mathbb{R}^-$ for $\Lambda<0$, can be rearranged into the well-known form of the Lambert equation
\begin{equation}
    y e^y = x,
    \label{eq: lambert eq}
\end{equation}
where we defined
\begin{equation}
    x\equiv \frac{\Lambda r_0^2}{\kappa^2 Q_1^2}e^{-\frac{M}{\kappa^2 Q_1^2}}.
\end{equation}
Given that $x<0$ holds by definition and $x,\,y \in \mathbb{R}$, the expression \eqref{eq: lambert eq} admits solutions only for $x\ge-e^{-1}$, and the cosmological constant is bounded from below by
\begin{equation}
    \Lambda \ge -\frac{\kappa^2 Q_1^2}{r_0^2}e^{\frac{M}{\kappa^2 Q_1^2}-1}.
\end{equation}
For $-e^{-1}\le x < 0$ the solutions for $y$ are displayed by the branches $y=W_0(x)$ and $y=W_{-1}(x)$, with $ W_{-1}(x)\le W_0(x)$. The radii of the inner and outer horizons are therefore given by
\begin{align}
    r^{in}_H(\Lambda)&=\sqrt{\frac{\kappa^2 Q_1^2}{\Lambda}W_{0}\leri{\frac{r_0^2e^{-\frac{M}{\kappa^2 Q_1^2}}}{\kappa^2 Q_1^2}\Lambda }}\\
    r^{out}_H(\Lambda)&=\sqrt{\frac{\kappa^2 Q_1^2}{\Lambda}W_{-1}\leri{\frac{r_0^2e^{-\frac{M}{\kappa^2 Q_1^2}}}{\kappa^2 Q_1^2}\Lambda }},
\end{align}
and they coincide for $\Lambda=\Lambda_{min}=-\frac{\kappa^2 Q_1^2}{r_0^2}e^{\frac{M}{\kappa^2 Q_1^2}-1}$, where $r^{in}_H=r^{out}_H=r_0\sqrt{e^{1-\frac{M}{\kappa^2 Q_1^2}}}$.
\\For $\Lambda\in(\Lambda_{min},0)$ the inner horizon is a monotonically decreasing function of $\Lambda$, and it reaches its minimum in $\Lambda_{max}=0$, where
\begin{equation}
    r_H^{in}=r_0\sqrt{e^{-\frac{M}{\kappa^2 Q_1^2}}}.
    \label{eq: inner horizon min}
\end{equation}
Then, taking the limit $Q_1\to\infty$ in \eqref{eq: inner horizon min}, we see that the fiducial length scale $r_0$ can be identified with the minimal value of the inner horizon in the asymptotic case $\Lambda\to 0^-$.
\\The outer horizon is instead a monotonically increasing function, which diverges for $\Lambda\to 0^-$
\bibliographystyle{JHEP}
\bibliography{references}
\end{document}